\documentclass[
    journal=apchd5,
    manuscript=article,
]{achemso}
\SectionNumbersOff

\usepackage{chemformula} % Formula subscripts using \ch{}
\usepackage[T1]{fontenc} % Use modern font encodings

\usepackage[colorlinks,linkcolor=blue,citecolor=blue]{hyperref}

\usepackage{color}

\usepackage{mathtools}
\usepackage{extarrows}
\usepackage{bm}% bold math

\usepackage{graphicx}

\usepackage{amsmath}
\usepackage{amssymb}

\author{Qingyi Zhou}
\email{qzhou75@wisc.edu}
\affiliation{
Department of Electrical and Computer Engineering, University of Wisconsin-Madison, Madison, WI 53706, USA
}

\author{S. Ali Hassani Gangaraj}
\affiliation{
Optical Physics Division, Corning Research and Development Corporation, Sullivan Park, Corning, NY 14831, USA
}

\author{Ming Zhou}
\affiliation{
Department of Electrical Engineering, Stanford University, Stanford, CA 94305, USA
}

\author{Zongfu Yu}
\affiliation{
Department of Electrical and Computer Engineering, University of Wisconsin-Madison, Madison, WI 53706, USA
}

\title[An \textsf{achemso} demo]
  {Simulating quantum emitters in arbitrary photonic environments using FDTD: beyond the semi-classical regime}

\abbreviations{IR,NMR,UV}

\begin{document}

\begin{abstract}
We propose a numerical algorithm that integrates quantum two-level systems into the finite-difference time-domain framework for simulating quantum emitters in arbitrary 3D photonic environments. Conventional semi-classical methods struggle with these systems due to spurious self-interactions that arise when a two-level system is driven by its own radiation field. We address this issue by determining the correct electric field for driving the two-level system, as well as the current source used in finite-difference time-domain for modeling photon emission. Our method, focusing on single-excitation states, employs a total field-incident field technique to eliminate self-interactions, enabling precise simulations of photon emission and scattering. The algorithm also successfully models complex phenomena such as resonant energy transfer, superradiance, and vacuum Rabi splitting. This powerful computational tool is expected to substantially advance research in nanophotonics, quantum physics, and beyond. 
\end{abstract}

\section{Introduction}
Classical electrodynamics, governed by Maxwell's equations, has played an important role in understanding and designing electromagnetic devices. 
This foundational theory has influenced a variety of scientific and technological fields. 
The most popular method for numerically solving Maxwell's equations is the finite-difference time-domain (FDTD) method \cite{Taflove_FDTD_textbook}, with efficient and mature solvers that are widely available \cite{Tidy3D, MEEP_2010}. 
As the frontiers of science push deeper into the quantum realm, advances in quantum physics \cite{quantum_dot_in_waveguide_NC_2015, quantum_nonlinear_optics_Lukin_2014} and chemistry \cite{polariton_chem_review_Zhou, polariton_chem_review_Mandel, Biswas_BIC_2024} are driving the demand for a more profound understanding of light-matter interactions at the quantum level, and the limitations of Maxwell's equations become evident. 
The dynamics of multiple quantum two-level systems (TLSs) interacting within complex photonic environment \cite{fidelity_exponential_improvement, Xuewen_PRL, waveguide_QED_review, Englund_PhC_cavity, ying_extend_range, chiral_atom_2016, super_Coulombic_hyperbolic_2017, diamond_waveguide_2016, Kejie_photon_photon_nonlinearity}, challenge existing simulation techniques. 
While analytical solutions are feasible for simple photonic environments like single-mode optical cavities \cite{Jaynes_1963, Tavis_1968, cavity_QED_tutorial_2021} or single-mode waveguides \cite{Shen_waveguide_single_photon_2005, Shen_waveguide_input_output_2010}, it's difficult to generalize these solutions to arbitrary photonic environment, and researchers still need to rely on numerical techniques. 

Currently, the simulation techniques used to tackle this issue can be divided into two categories: semi-classical methods, and approaches based on master equation \cite{Ficek_quantum_textbook, Kimble_1D_Green_func_2017, elements_quantum_optics_textbook_2007}. 
The semi-classical methods, such as Maxwell-Bloch \cite{ultrafast_pulse_1995, mbsolve_2021} or Maxwell-Schr$\ddot{\text{o}}$dinger equations \cite{Sha_Maxwell_Schrodinger, Maxwell_Schrodinger_control_pulse_2015, multiscale_Maxwell_Schrodinger_2009}, are limited by their reliance on classical fields, and often fail to correctly account for incoherent processes \cite{Li_comparison_2019}. They also require careful treatment of self-interaction, which is often overlooked in previous works \cite{Self_interaction_Maxwell_Liouville, self_consistent_Lorentz_Hughes_2017}. 
%On the other hand, for master equation approaches the photon degrees of freedom is often traced out. These approaches rely on calculating dyadic Green's functions and become computationally expensive for large number of quantum elements. Moreover, employing the Born-Markov approximation leads to inaccurate results when memory effects become important \cite{Milonni_retardation_1974, observe_non_Markovian_2011, light_cone_spread_2012}. 
The above issues have made it quite difficult to accurately simulate multiple TLSs in arbitrary environment. Currently, a general algorithm that has been thoroughly tested is still lacking, which hinders research in quantum physics and nanophotonics. 

In this paper, we aim to address the above issues and provide a simulation technique that's available for use. We propose an algorithm based on 3D FDTD due to the fact that FDTD is both versatile and highly efficient. To incorporate quantum two-level systems (TLSs) into FDTD, the problem is first simplified by focusing on single-excitation states. 
We rigorously analyze the dynamics of TLSs driven by the electric field, as well as how TLSs couple back to Maxwell's equations through radiation emission. 
To avoid unwanted self-interaction, which is essential for accurate results, we propose a total field-incident field (TF-IF) technique to exclude the primary radiation field from driving the TLS. 
We first validate our approach through benchmark examples involving one TLS, demonstrating that our method accurately computes the spontaneous emission rates and scattering cross sections, which is not possible without mitigating self-interaction effects. 
Further, we extend our simulations to systems involving $N \geqslant 2$ TLSs, exploring phenomena such as excitation transport between TLSs, superradiance in TLS arrays, as well as vacuum Rabi splitting when TLSs are strongly coupled to a ring resonator. 
These examples demonstrate the capability of our proposed algorithm to simulate dynamics of multiple TLSs placed within complex 3D photonic environments, which, to the best of our knowledge, has never been conducted successfully before. 
The implementation of this algorithm has been realized in CUDA C++, with the code made publicly available on \href{https://github.com/zhouqingyi616/FDTD_w_TLS}{GitHub}. Currently, we are working on integrating this algorithm into Tidy3d \cite{Tidy3D} simulation platform to ensure its availability in the near future, with the hope that it will serve as a valuable tool for researchers across various fields. 

\section{Results}

\subsection{Formulation of the proposed FDTD}
% Help people understand the field, review the popular methods people use, and introduce a bit about the limitation of current methods. 
The optical properties of hybrid systems that combine quantum emitters (such as atoms or quantum dots) with complex photonic environment are of much current interest. 
% It is possible to derive analytical solution for very simple structures such as single-mode optical cavity \cite{Jaynes_1963} or waveguide \cite{Shen_waveguide_single_photon_2005, Shen_waveguide_input_output_2010} under certain approximations. 
When the problem involves multiple TLSs inside a complicated environment (for example, a multi-mode cavity \cite{route_multimode_cavity_2014, beyond_strong_coupling_multimode_2015, multimode_lossy_resonator_2023, intermolecular_in_cavity_2021}, or photonic crystal \cite{ying_extend_range, inhibited_decay_Yablonovitch_1987, Kimble_2D_PhC_slab_2019, dipole_radiation_Dirac_PhC_2020, RDDI_PhC_slab_cavity_2012, long_range_conelike_bath_2018}), it becomes necessary to incorporate TLSs into full-wave electromagnetic simulation since analytical solution might not be available. 
Many researchers have been working on this topic, and have developed semi-classical methods \cite{ultrafast_pulse_1995,  mbsolve_2021, Sha_Maxwell_Schrodinger, Maxwell_Schrodinger_control_pulse_2015, multiscale_Maxwell_Schrodinger_2009, Self_interaction_Maxwell_Liouville, self_consistent_Lorentz_Hughes_2017, time_discretize_Maxwell_Bloch_2003, FDTD_ultrashort_pulse_OL_2003, nonlinear_Maxwell_Schrodinger_2009, FDTD_TLS_PhC_cavity_Japan_2011, Maxwell_Schrodinger_Ryu_2016, Maxwell_Bloch_PCSEL_Hughes_2017, chiral_modeling_2023} by combining classical Maxwell's equations with an extra set of equations that can describe the TLSs' dynamics. % The interaction between different TLSs need to be calculated in advance before start solving the master equation. The calculation of dipole-dipole interaction relies on calculating the dyadic Green's function, which leads to overhead and will become expensive when the number of involved TLSs becomes large. 
%Our proposed method is similar to the semi-classical methods mentioned above. Before introducing the implementation details, we will first answer two vital questions: (1) what is the electric field value that should be used to drive the TLS; (2) what is the source term that should be used in Maxwell's equations when considering radiation of the TLS. As will be shown, the answers to these questions are crucial for incorporating TLS dynamics into Maxwell's equations correctly. 

% \subsection{Driving term related to TLS dynamics}
To correctly simulate a hybrid system containing both TLSs and photonic environment, it is required to understand how the electromagnetic fields affect the dynamics of TLSs. 
Without loss of generality, we first focus on the dynamics of one TLS. 
The resonance frequency of the TLS is denoted as $\omega_{0}$, and the corresponding resonance wavelength is represented by $\lambda_{0}$. Its spontaneous emission rate inside vacuum is denoted as $\Gamma_\text{vac}$. These notations are consistently used throughout the paper. 
% To make our discussions general, instead of limiting ourselves to a certain set of equations, we use a more generalized notation $\vec{u}(t)$ to describe the status of TLS. When a TLS is driven by electric field, the governing ordinary equations can be summarized as 
% \begin{equation}
% \begin{aligned}
%     \frac{d}{dt}\vec{u}(t) 
%     & = \textcolor{red}{ \{ \text{Decay term} \} } + \textcolor{blue}{ \{ \text{Driving term} \} } \\
%     & \xlongequal{\Delta} \textcolor{red}{ \Gamma \mathcal{L}(\vec{u}) } + \textcolor{blue}{ f(\vec{u}) \cdot \vec{E} },
%     \label{eq:du_dt}
% \end{aligned}
% \end{equation}
% where $\Gamma$ denotes the decay rate, and $\vec{E}$ denotes the electric field value that's used to drive the TLS. 
Physically, the electric fields inside the simulation domain can be divided into $3$ parts: $\vec{E}_\text{tot} = \vec{E}_\text{inc} + \vec{E}_\text{rad} + \vec{E}_\text{ref}$, where $\vec{E}_\text{inc}$ represents the externally applied field in the presence of photonic structure; $\vec{E}_\text{rad}$ represents the primary radiation field emitted by the TLS, before hitting any photonic structures; $\vec{E}_\text{ref}$ represents the radiation field reflected after hitting photonic structure. 
% Fig.~\ref{fig:fig1}(b) provides a high-level illustration of the hybrid system, together with all electric field components. 
Both $\vec{E}_\text{inc}$ and $\vec{E}_\text{ref}$ should be included in the driving term $\vec{E}$ with no doubt. However, the $\vec{E}_\text{rad}$ field produced by a dipole source is nonzero at its own position. Therefore, using the $\vec{E}_\text{tot}$ field obtained from FDTD results in the TLS being driven by the radiation field produced by itself, which is incorrect physically. 
% This leads to two effects: first, energy is carried away from TLS, causing it to decay at rate $\Gamma_\text{vac}$; second, this self-interaction causes the TLS's resonance frequency to shift. This phenomenon, which can be understood as a numerical analogy of Lamb shift \cite{Bethe_Lamb_1947}, has been observed in literatures \cite{Self_interaction_Maxwell_Liouville, self_consistent_Lorentz_Hughes_2017}.  
\begin{figure*}[]
\centering\includegraphics[width=1.0\textwidth]{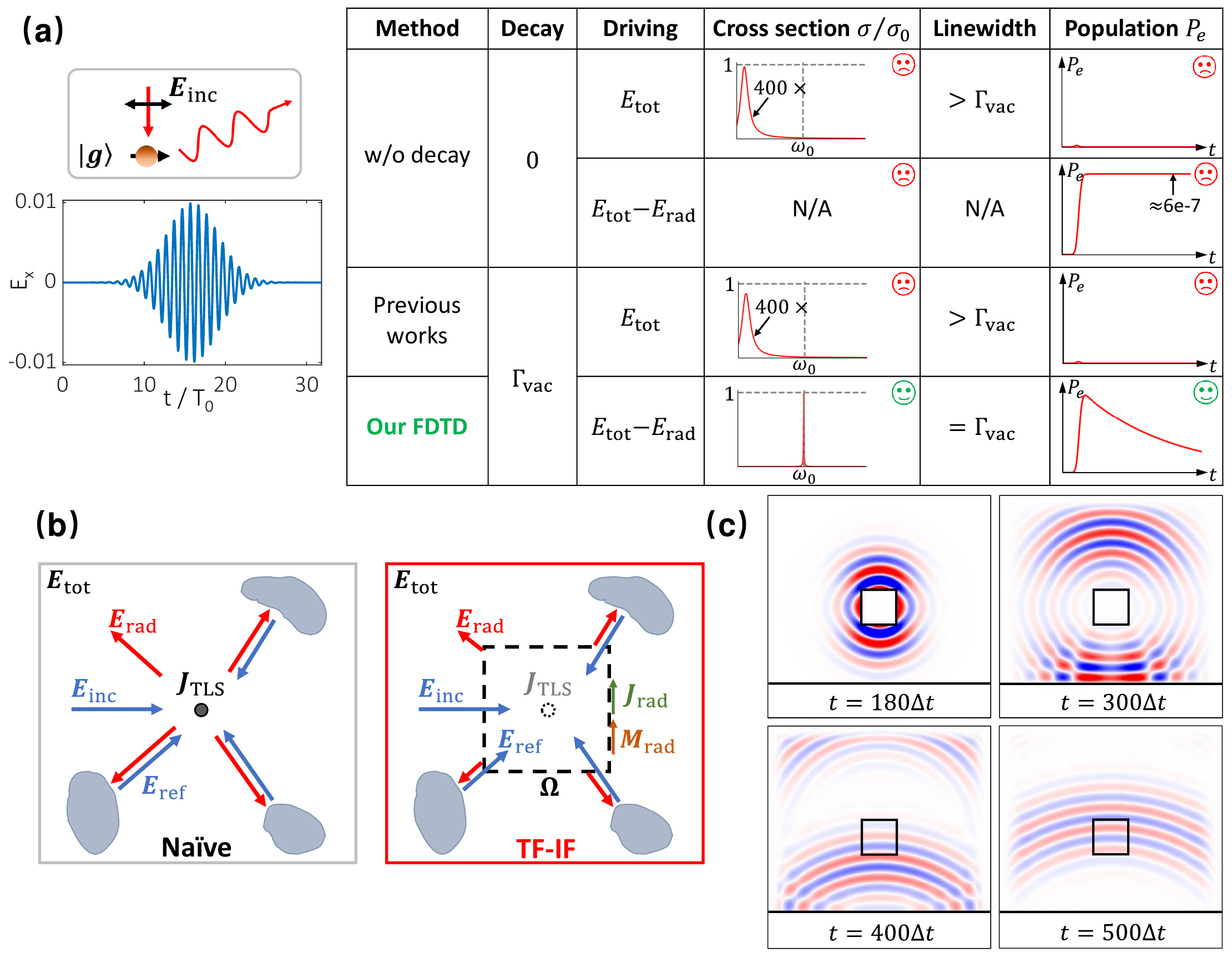}
\caption{The exclusion of primary radiation field produced by TLS. (a) Light scattering of one TLS inside vacuum. The inset shows the time profile of the incident Gaussian pulse. The table presents simulation results for 4 scenarios. Without excluding the radiation field $\vec{E}_\text{rad}$, the scattering cross section $\sigma(\omega)$ cannot be predicted correctly. Our FDTD can get rid of the self-interaction caused by primary radiation, thus providing accurate results. 
(b) Comparison of the TF-IF technique with the na\"{i}ve approach. When using TF-IF, the primary radiation $\vec{E}_\text{rad}$ only exists outside region $\Omega$, making sure that unwanted self-interactions are eliminated. (c) The effect of TF-IF technique. An oscillating dipole in $x$ direction is placed at the center of 3D domain. Nonzero $E_{x}$ field is produced outside region $\Omega$ (marked as a square box). After the radiation field has been reflected by a PEC mirror, it can enter $\Omega$ and drive the oscillating dipole. Here $\Delta t$ denotes the time step. }
\label{fig:fig1}
\end{figure*}
To verify that $\vec{E}_\text{rad}$ should not be included in the driving term, we now consider a simple example, where an incident pulse is scattered by a single TLS placed inside vacuum. It is well known that under weak-excitation limit, a TLS acts like an oscillating dipole \cite{Steck_notes_2007}. Its scattering cross section $\sigma(\omega)$ should follow a Lorentzian line shape, with maximum value $\sigma_{0}=\frac{3\lambda_{0}^{2}}{2\pi}$ and an FWHM of $\Gamma_\text{vac}$ \cite{perfect_reflection_PRL_2008, Liu_photon_scatter_OL_2016, Ming_large_cross_section_PRL_2015, Liu_enhance_cross_section_PRA_2017, maximize_scattering_tutorial_2014}. 
The simulation setup is shown schematically in Fig.~\ref{fig:fig1}(a). A Gaussian pulse with $x$-polarization serves as the incident wave, whose time-profile is plotted in the inset. The TLS, whose dipole moment $|\langle g | \hat{\vec{d}}| e \rangle | = 0.02$, stays at its ground state $| g \rangle$ at $t=0$. 
The TLS is simulated using Bloch equation (see Methods), and four different scenarios are checked: 
\begin{enumerate}
    \item Decay rate $\Gamma=0$, $\vec{E}_\text{rad}$ not excluded;
    \item Decay rate $\Gamma=0$, $\vec{E}_\text{rad}$ excluded;
    \item Decay rate $\Gamma=\Gamma_\text{vac}$, $\vec{E}_\text{rad}$ not excluded (corresponds to most existing works);
    \item Decay rate $\Gamma=\Gamma_\text{vac}$, $\vec{E}_\text{rad}$ excluded (corresponds to our proposed FDTD).
\end{enumerate}
We surround the TLS with a power monitor to calculate the power of scattered electromagnetic field, then divide it by the incident light intensity to calculate the scattering cross section $\sigma(\omega)$. 
The results are summarized in the table shown in Fig.~\ref{fig:fig1}(a). Based on the first and the third rows, it can be concluded that if $\vec{E}_\text{rad}$ is not excluded, the scattering cross section $\sigma$ is much smaller than $\sigma_{0}$ (noting that the $\sigma(\omega)$ curves are exaggerated by $400\times$). Also the resonance frequency shifts away from $\omega_{0}$. 
On the other hand, the second row shows that by setting $\Gamma=0$ and excluding $\vec{E}_\text{rad}$, the TLS's population $P_{e}(t)$ does not decay, due to the lack of a decaying mechanism. 
As a comparison, our FDTD correctly recovers the position, linewidth, as well as the maximum value of the $\sigma(\omega)$ peak. The exclusion of primary radiation $\vec{E}_\text{rad}$ is essential for avoiding spurious self-interaction. 
%Note that some researchers have tried to mimic TLS by modifying the dielectric constant $\epsilon_{r}(\omega)$ of the grid point \cite{self_consistent_Lorentz_Hughes_2017, FDTD_TLS_PhC_cavity_Japan_2011}. Though the derivation of such dielectric function is well-known \cite{Boyd_nonlinear}, it has been pointed out that this method does not exclude the self-interaction \cite{Li_comparison_2019}, making it incorrect when simulating isolated TLSs. 
%The third effect can be seen from Fig.~\ref{fig:fig1}(a): when driven by the primary radiation field at its own location, the TLS's excited probability will oscillate strongly (see the curve labeled with ``FDTD w/o TF-IF''). This non-physical oscillation is strong because the radiation field produced at TLS's own location can become very large, especially when the gris size $\Delta x$ is small. 
% Solution: using TF-IF
In the proposed FDTD we enclose the TLS with an imaginary square domain $\Omega$. 
We hope that the primary radiation field of this TLS only exists outside $\Omega$, thus the field $\vec{E}$ sampled inside $\Omega$ can be used to drive TLS. 
To achieve this, we utilize the surface equivalence principle \cite{Taflove_FDTD_textbook, EM_theory_Jin_2015}: instead of directly using the TLS as a dipole source, fictitious surface current densities are put on region boundary $\partial \Omega$ (as shown in Fig.~\ref{fig:fig1}(b)), ensuring that the same radiation field is excited outside $\Omega$. 
We refer to this modification as total field-incident field (TF-IF) technique, which divides the simulation domain into different regions: within $\Omega$ only the electromagnetic fields incident from outside exist; outside $\Omega$ the fields can be understood as ``total field'' $\vec{E}_\text{tot}$, including the primary radiation from the TLS. 
The modification and its name have been inspired by the total field-scattered ﬁeld (TF-SF) technique, commonly used in FDTD simulations to generate incident wave \cite{Taflove_FDTD_textbook}. 
% Consequence
The consequence of this TF-IF modification is depicted in Fig.~\ref{fig:fig1}(c). Here we consider a simple 3D FDTD simulation with a dipole source $3\lambda_{0}$ above a PEC mirror. The dipole produces a pulse-like radiation field, and it is observed that the primary radiation field exists only outside region $\Omega$ (marked by a square box). After the primary radiation field hits the mirror (marked by a black line) and gets reflected back, the reflected field $\vec{E}_\text{ref}$ enters $\Omega$ and re-excites the TLS. Therefore, the electric field sampled inside $\Omega$ can be used directly as a driving term. 

\begin{figure*}[]
\centering\includegraphics[width=1\textwidth]{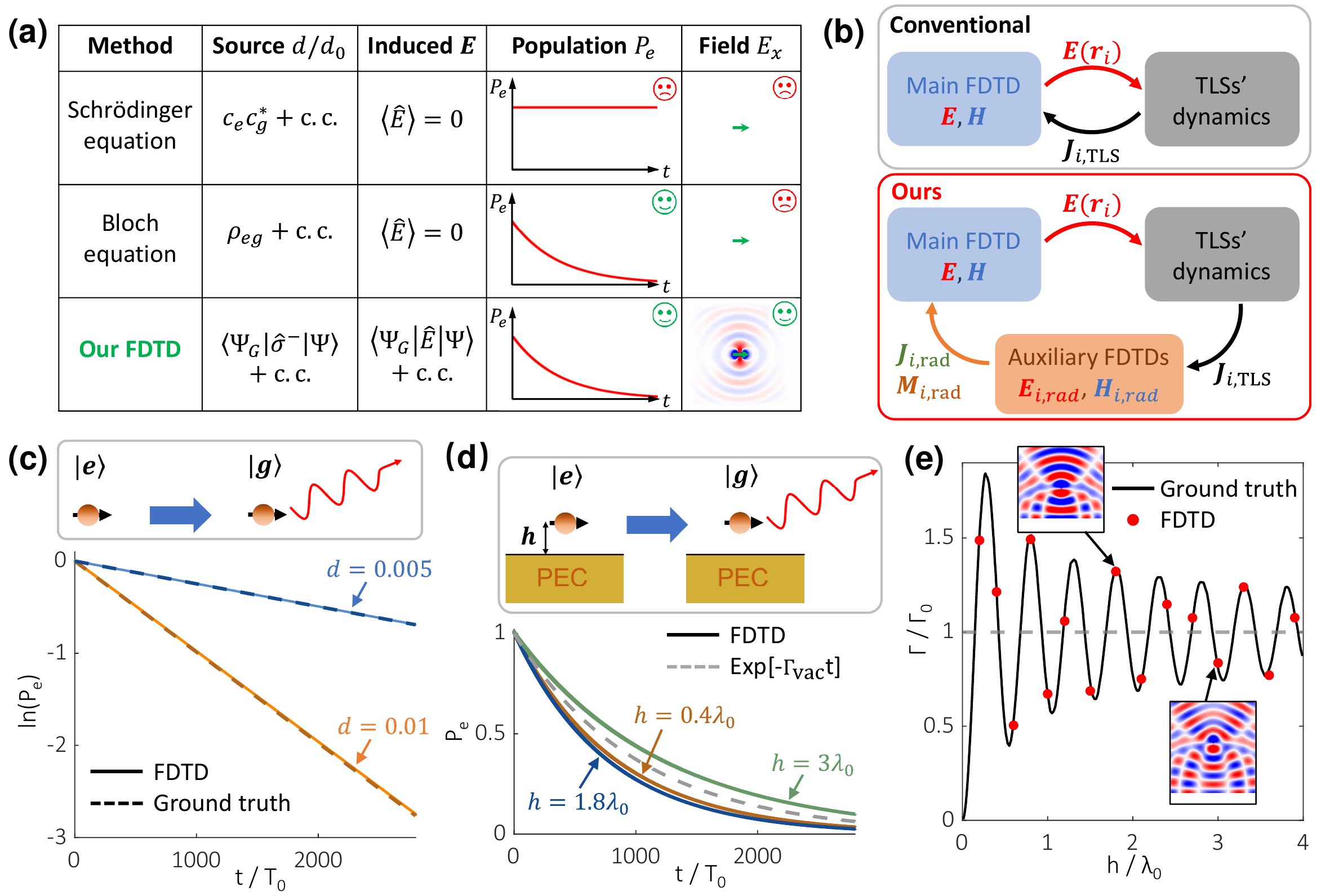}
\caption{Using the correct current source makes it possible to simulate photon emission process. 
(a) Spontaneous emission of one TLS inside vacuum. Both Bloch equation and our FDTD recover an exponential decay in population. However the two semi-classical methods fail to produce nonzero fields since $\langle \hat{E} \rangle = 0$.
(b) Flowchart of the proposed FDTD. The primary radiation of each TLS is calculated using an auxiliary FDTD. On the contrary, conventional methods do not exclude self-interaction.
(c) Spontaneous emission of one TLS inside vacuum. 
(d) Spontaneous emission of one TLS placed above a PEC mirror. Time-evolution of $P_{e}$ have been plotted for different distances $h/\lambda_{0}\in \{ 0.4, 1.8, 3 \}$. 
(e) The relationship between decay rate $\Gamma$ and distance $h$. Results obtained by FDTD match perfectly with the ground truth. The insets show the corresponding $H_{z}$ field distributions. Note that the simulations involved in (c)(d)(e) are carried out in 2D domain. }
\label{fig:fig2}
\end{figure*}

% Compare
Our method is similar to the one proposed in Ref.~\citenum{Self_interaction_Maxwell_Liouville}. Yet here our $\Omega$ region can be much smaller (as small as $3\times 3\times 3$ grid points) compared with the square region used in Ref.~\citenum{Self_interaction_Maxwell_Liouville}. This flexibility is due to the radiation field being numerically calculated using an auxiliary FDTD (see Supporting Information Note S3 for details). This advantage enables us to model multiple TLSs that are placed very close (as close as the grid resolution $\Delta x$) to each other, which will be demonstrated in the following sections. In contrast, simulations involving multiple TLSs are not presented in Ref.~\citenum{Self_interaction_Maxwell_Liouville}. 
% \subsection{Current source related to TLS radiation}
More specifically, we consider a more general system containing $N$ TLSs. The $i$-th TLS is located at position $\vec{r}_{i}$ ($i=1,2,...,N$). We limit ourselves to the  single-excitation state
\begin{equation}
    | \Psi(t) \rangle = \sum_{i} b_{i}(t) |e_{i},0\rangle + \sum_{\vec{k}, \lambda} c_{\vec{k}\lambda}(t) | g,1_{\vec{k}\lambda}\rangle, 
    \label{eq:state}
\end{equation}
where $b_{i}(t)$ represents the $i$-th TLS's excitation amplitude, and $c_{\vec{k}\lambda}(t)$ corresponds to the single-photon state with wave vector $\vec{k}$ and polarization $\lambda$. Such single-excitation states, although simple, capture all phenomena in linear optics regime, and contain rich physics \cite{fidelity_exponential_improvement, Shen_waveguide_single_photon_2005, Kimble_1D_Green_func_2017, Liu_enhance_cross_section_PRA_2017, collective_Lamb_shift_Scully_2009, RDDI_2002, strong_long_range_ying_2022, Ali_nonreciprocal_2022, Scully_timing_2006}. The ground state of this system is denoted as $| \Psi_{G} \rangle = |g,0\rangle$. 
Instead of simulating the expectation value $\langle \vec{E} \rangle$, we now use FDTD to simulate the time-evolution of the electric field $\vec{E}(\vec{r}, t)$, defined as $\vec{E}(\vec{r}, t) = \langle \Psi(t) | \hat{\vec{E}}(\vec{r}) | \Psi_{G} \rangle + \langle \Psi_{G} | \hat{\vec{E}}(\vec{r}) | \Psi(t) \rangle$, which is nonzero for single-photon state, making it an ideal choice for simulating photon emission \cite{Scully_single_photon}. 
It can be proved that the current source $\vec{J}_{\text{TLS}}(\vec{r})$ that should be introduced in FDTD now becomes (see Supporting Information Note S1)
\begin{equation}
% \begin{aligned}
    \vec{J}_{\text{TLS}}(\vec{r}) 
    = 2\omega_{0} \sum_{i} \vec{d}_{i} \cdot \text{Im}(b_{i}) \cdot \delta(\vec{r} - \vec{r}_{i}), 
    \label{eq:J_TLS}
% \end{aligned}
% \textcolor{red}{\text{~~(proposed)}}
\end{equation}
where $\text{Im}(b_{i})$ stands for the imaginary part of $b_{i}(t)$. 
To justify the above choices, we examine the spontaneous decay of an excited TLS inside vacuum. In Fig.~\ref{fig:fig2}(a) we compare our FDTD with two baseline semi-classical simulation techniques, namely, the Maxwell-Schr$\ddot{\text{o}}$dinger equations and the Maxwell-Bloch equations (implementation details can be found in Methods). 
By plotting the time-evolution of the excited probability $P_{e}$, it is evident that the Maxwell-Schr$\ddot{\text{o}}$dinger equation does not lead to spontaneous decay, and the TLS remains at the excited state, which is consistent with previous observations \cite{Sha_Maxwell_Schrodinger, Li_comparison_2019, Ehrenfest_R_2019}. Both Maxwell-Bloch equation and our FDTD leads to an exponential decay as $\exp(-\Gamma_\text{vac}t)$. 
On the other hand, neither the Schr$\ddot{\text{o}}$dinger equation nor the Bloch equation produces a nonzero $\vec{E}$ field, while our proposed FDTD can accurately simulate the photon emission process. 

% Incorrect decay rate w/ structure
% The above problem becomes even more serious if we consider the TLS's decay rate in the presence of photonic structures. It has been well known that the decay rate of a quantum emitter will change when it's put inside a non-trivial photonic environment. For example, when the emitter is weakly coupled to a cavity mode, its decay rate gets enhanced. However, existing semi-classical simulation techniques cannot reproduce this modification of decay rate correctly. As we've stated, for a spontaneously decaying TLS, these semi-classical methods do not produce nonzero electromagnetic fields. Therefore, when utilizing Maxwell-Schr$\ddot{\text{o}}$dinger equations the excited TLS does not decay at all; when utilizing Maxwell-Bloch equations, the decay rate simply follows the phenomenological parameter contained in the Bloch equation, which does not change accordingly when the photonic environment changes. 

% \subsection{Process}
Finally, by combining the two points mentioned previously, we briefly summarize the entire process of the proposed FDTD algorithm (details can be found in Methods). 
We restrict our analysis to single-excitation quantum states defined in eq.~(\ref{eq:state}). It can be proved that the time-evolution of $\vec{E}$ and $\vec{H}$ fields still follow classical Maxwell's equations (Supporting Information Note S1), and thus can be simulated using FDTD without any difficulty. 
On the other hand, the time-evolution of the $i$-th TLS's excitation amplitude $b_{i}$ follows (Supporting Information Note S1)
\begin{equation}
    \frac{db_{i}}{dt} = (-i\omega_{0} - \frac{\Gamma_{\text{vac}}}{2}) b_{i} + i\frac{\vec{d}_{i} \cdot \vec{E}(\vec{r}_{i}, t)}{\hbar}. 
    \label{eq:db_dt}
\end{equation}
As previously stated, $\vec{E}(\vec{r}_{i})$ is sampled inside region $\Omega_{i}$ and the primary radiation has been excluded. 
The above two equations, namely, eq.~(\ref{eq:J_TLS})(\ref{eq:db_dt}), form the core of our modified FDTD algorithm. 
Before starting the FDTD simulation, we first initialize a 3D domain containing all photonic structures as well as all $\Omega_{i}$ regions. The FDTD carried out in this domain is referred to as the ``main FDTD''. 
Additionally, each TLS with index $i$ requires an auxiliary FDTD to calculate the primary radiation fields $(E_{i, \text{rad}}, H_{i, \text{rad}})$. These $N$ FDTD simulations, filled with homogeneous media, are termed ``auxiliary FDTD'' and are kept small to minimize computational overhead (see Supporting Information Note S3 for more details). 
At each time step, our proposed FDTD algorithm comprises 3 parts: 
\begin{enumerate}
    \item Update the main FDTD for one step, based on the current sources $(\vec{J}_{i, \text{rad}}, \vec{M}_{i, \text{rad}})$ provided by auxiliary FDTDs.
    \item Update all $N$ TLSs for one step, using the $\vec{E}(\vec{r}_{i})$ field sampled from the main FDTD.
    \item Update all $N$ auxiliary FDTDs for one step, treating the TLSs as dipole sources.
\end{enumerate}
The above procedure has been illustrated in Fig.~\ref{fig:fig2}(b). Based on the surface equivalence principle \cite{Taflove_FDTD_textbook, EM_theory_Jin_2015}, the fictitious current sources $(\vec{J}_{i, \text{rad}}, \vec{M}_{i, \text{rad}})$ used in the main FDTD are related to the primary radiation fields $(\vec{E}_{i, \text{rad}}, \vec{H}_{i, \text{rad}})$ by: $\vec{J}_{i, \text{rad}}=\hat{n} \times \vec{H}_{i, \text{rad}}$, $\vec{M}_{i, \text{rad}}=-\hat{n} \times \vec{E}_{i, \text{rad}}$ (here $\hat{n}$ denotes the normal vector of region surface $\partial \Omega_{i}$). For more implementation details, please refer to Supporting Information Note S3. 

\subsection{Spontaneous emission: $N=1$ case}
In this part, we provide several benchmark examples involving the spontaneous emission of $N=1$ TLS. 
The simplest case would be a TLS spontaneously decaying inside vacuum. Here we test two TLSs, with different dipole moments $d\in \{ 0.005, 0.01\}$. The time-evolution of excited probability $P_{e}$ during the decay process is plotted in Fig.~\ref{fig:fig2}(c). It can be concluded that the proposed FDTD algorithm can reproduce the exponential decay $P_{e}(t) \sim \exp(-\Gamma_{\text{vac}}t)$. 

Next, to prove that our algorithm can capture the influence of photonic environment, we consider a TLS with dipole moment $d_{x}=0.01$, located off a PEC mirror. The setup is shown in Fig.~\ref{fig:fig2}(d), with the distance between TLS and mirror denoted as $h$. In Fig.~\ref{fig:fig2}(d) we've plotted the time-evolution of $P_{e}$ for three different cases, $h=0.4\lambda_{0}$, $1.8\lambda_{0}$, $3.0\lambda_{0}$. 
To verify that our FDTD can predict modified decay rates correctly, we further run multiple FDTD simulations for different heights $h$, ranging from $0\sim 4\lambda_{0}$. The corresponding decay rates $\Gamma$, obtained through fitting $P_{e}(t)$ with $\exp(-\Gamma t)$, are compared with analytical results in Fig.~\ref{fig:fig2}(e). As can be seen from the comparison, in all test cases our FDTD can predict the modified decay rate perfectly. By limiting ourselves to single-excitation states, we have successfully simulated the dynamics of the TLS when it gets re-excited by the reflected photon, leading to the correct modified decay rate.

% Note that it's not possible to arrive at the above results if semi-classical method is used to simulate a TLS starting from excited state, since a TLS with zero coherence cannot produce nonzero fields in FDTD. In that case, the spontaneous decay rate won't be affected by photonic structure, which is why many researchers tend to introduce a phenomenological parameter to compensate for the modified decay rate \cite{Xuewen_PRL, mbsolve_2021, multiscale_Maxwell_Schrodinger_2009, FDTD_ultrashort_pulse_OL_2003, FDTD_TLS_PhC_cavity_Japan_2011, chiral_modeling_2023}. In previous works, researchers often intentionally circumvent this issue, either by setting the initial state of the TLS as a superposition of the ground state and the excited state \cite{FDTD_TLS_PhC_cavity_Japan_2011, RDDI_Li_2018, spontaneous_decay_zero_point_energy_Li_2018}, or by using an incident pulse at the beginning to excite the TLS \cite{ultrafast_pulse_1995, Sha_Maxwell_Schrodinger, Maxwell_Schrodinger_control_pulse_2015,  multiscale_Maxwell_Schrodinger_2009, chiral_modeling_2023}. Our proposed FDTD can resolve this issue simply by restricting ourselves to single-excitation states. In this way, we have successfully simulated the dynamics of the TLS when it gets re-excited by the reflected photon, leading to the correct modified decay rate. 

\subsection{Dipole-dipole interaction: $N=2$ case}
Understanding photon exchange between TLSs (often referred to as resonant energy transfer) is fundamental to the development of modern quantum technologies \cite{RET_review_2019, RDDI_QED_approach_2004, RDDI_Rydberg_atoms_2014}. This type of energy transfer also plays a crucial role in a variety of biological and chemical systems \cite{FET_photosynthesis, long_range_RET_2003, FRET_quantum_dot_bioconjugates_2017, FRET_pitfalls_2019}. 
Given its importance, there have been substantial efforts to understand the underlying mechanisms of energy transport and to determine how these interactions can be controlled by engineering the photonic environment. 
\begin{figure*}[]
\centering\includegraphics[width=1\textwidth]{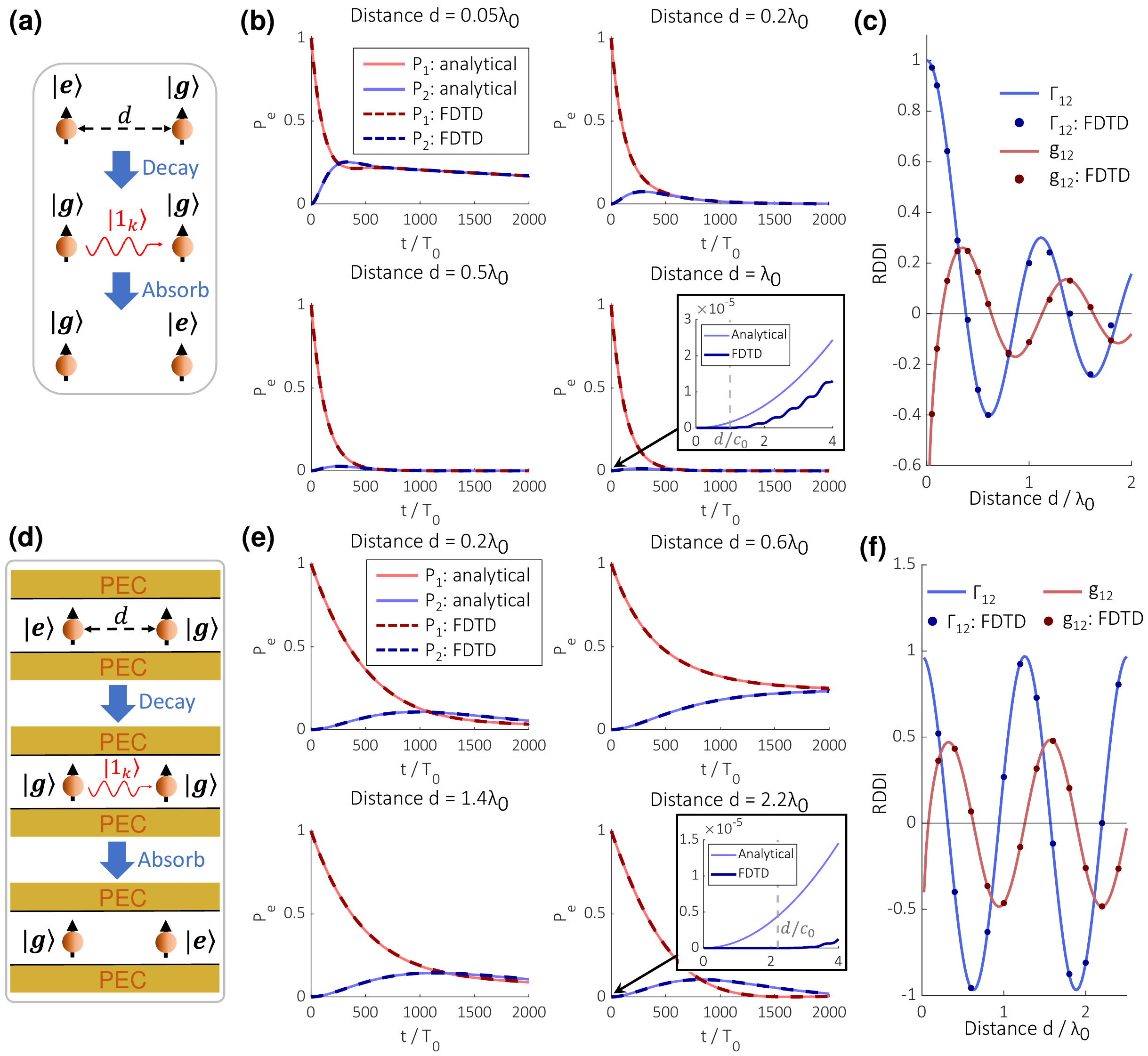}
\caption{2D FDTD simulations involving $N=2$ TLSs, focusing on excitation transport between an excited TLS and an unexcited one. (a) Illustration of 2 TLSs placed inside vacuum. The excited TLS emits a photon, which can then be absorbed by the unexcited TLS. (b) The time-evolution of excited probabilities $P_{1}(t)$ and $P_{2}(t)$. The inset shows that FDTD presents retardation effect and does not violate causality. (c) The dipole-dipole interaction strength (normalized by $\Gamma_{\text{vac}}$) extracted from FDTD simulation. (d) Illustration of 2 TLSs placed at the middle of a waveguide, formed by 2 PEC mirrors. (e) The time-evolution of excited probabilities $P_{1}(t)$ and $P_{2}(t)$. Similar to (b), the inset shows that FDTD presents retardation effect. (f) The dipole-dipole interaction strength (normalized by $\Gamma_{\text{vac}}$) extracted from FDTD simulation. The interaction strength oscillates when increasing distance $d$, because emitted photon is confined by the waveguide. }
\label{fig:fig3}
\end{figure*}
In this section, we simulate the excitation transport process between two TLSs using our FDTD algorithm. Consider two identical TLSs, where the $i$-th TLS ($i=1,2$), characterized by a dipole moment $\vec{d}_{i}$, is positioned at $\vec{r}_{i}$. Theoretically, the dipole-dipole interaction between these TLSs can be evaluated based on the dyadic Green's function $\stackrel{\leftrightarrow}{G}(\vec{r}_{i}, \vec{r}_{j})$:
\begin{align}
    \Gamma_{ij} &= \frac{2\omega_{0}^{2}}{\hbar \epsilon_{0} c_{0}^{2}} \vec{d}_{i} \cdot \text{Im}\stackrel{\leftrightarrow}{G}(\vec{r}_{i}, \vec{r}_{j}) \cdot \vec{d}_{j}, \nonumber \\
    g_{ij} &= \frac{\omega_{0}^{2}}{\hbar \epsilon_{0} c_{0}^{2}} \vec{d}_{i} \cdot \text{Re}\stackrel{\leftrightarrow}{G}(\vec{r}_{i}, \vec{r}_{j}) \cdot \vec{d}_{j},
\end{align}
where $\Gamma_{ij}$ represents the collective decay rate, and $g_{ij}$ represents the coherent coupling. 
At $t=0$ only the first TLS is at its excited state. After the first TLS emits a photon spontaneously, the second TLS can absorb this photon, leading to an increase of its excited probability. Based on the master equation \cite{Ficek_quantum_textbook, RDDI_2002, Ali_nonreciprocal_2022, RDDI_in_cavity_1997}, the excited probabilities of these two TLSs are
\begin{equation}
    P_{1}(t) = \frac{1}{4} \left[ e^{-(\Gamma_{11} + \Gamma_{12})t} + e^{-(\Gamma_{11} - \Gamma_{12})t} \right] + \frac{e^{-\Gamma_{11}t}}{2} \cos(2 g_{12} t),
    \label{eq:P1}
\end{equation}
\begin{equation}
    P_{2}(t) = \frac{1}{4} \left[ e^{-(\Gamma_{11} + \Gamma_{12})t} + e^{-(\Gamma_{11} - \Gamma_{12})t} \right] - \frac{e^{-\Gamma_{11}t}}{2} \cos(2 g_{12} t).
    \label{eq:P2}
\end{equation}

We now present the results of our FDTD simulations for two different 2D test cases, then compare with those analytical solutions obtained from the master equation. In the first scenario, two identical TLSs with resonance wavelength $\lambda_{0} = 1~\mu\text{m}$ are positioned in vacuum, as depicted in Fig.~\ref{fig:fig3}(a). In the second scenario, the same TLSs are placed inside a waveguide formed by two parallel PEC boards, as illustrated in Fig.~\ref{fig:fig3}(d). 
The separation distance between TLS 1 and TLS 2 is denoted as $d$. 
For the vacuum case, the time-evolution of excited probabilities $P_{1}(t)$ and $P_{2}(t)$ are depicted using dashed lines (see Fig.~\ref{fig:fig3}(b)). The corresponding analytical solutions are also plotted. We have explored four different inter-TLS distances, $d/\lambda_{0} \in \{ 0.05, 0.2, 0.5, 1\}$, and our FDTD results agree pretty well with analytical results eq.~(\ref{eq:P1})(\ref{eq:P2}) for all different distances. 
For the case where $d=\lambda_{0}$, an inset has been added to show a detailed view of $P_{2}(t)$ over the time interval $t\in \left[ 0, 4T_{0} \right]$. The $P_{2}(t)$ obtained from FDTD remains zero until $t_{r} = d / c_{0}$ (highlighted by a dashed gray line), due to the fact that electromagnetic wave travels with finite speed $c_{0}$ in FDTD. 
To verify the accuracy of our FDTD algorithm, we estimate the corresponding collective decay rate $\Gamma_{12}$ and the coherent coupling $g_{12}$, based on the $P_{1}(t)$, $P_{2}(t)$ curves obtained from FDTD simulations. The estimation is carried out through a curve fitting process based on the form of analytical solution eq.~(\ref{eq:P1})(\ref{eq:P2}). The comparison between the estimated values and ground truth is shown in Fig.~\ref{fig:fig3}(c). It can be concluded that FDTD maintains high accuracy across most test scenarios. %In contrast, although the semi-classical method introduced in \cite{RDDI_Li_2018} can capture the retardation effect, it does not accurately determine the dipole-dipole interaction strength. 

To prove that our FDTD handles the dipole-dipole interaction correctly with the presence of photonic structures, we introduce a second test case featuring a waveguide formed by two parallel PEC mirrors. The width of this waveguide is $w=0.8\lambda_{0}$, and the two TLSs are placed at the center of the PEC waveguide, as depicted in Fig.~\ref{fig:fig3}(d). Similar to the previous test case, we obtain the time-evolution of excited probabilities $P_{1}(t)$ and $P_{2}(t)$ from FDTD simulation across four different distances $d/\lambda_{0} \in \{ 0.2, 0.6, 1.4, 2.2\}$. In Fig.~\ref{fig:fig3}(e), the FDTD results are plotted using dashed lines, while the corresponding solutions obtained from master equation are also plotted for comparison. Once again, our FDTD agrees well with the analytical solutions under all tested conditions. Notably, for the $d=2.2\lambda_{0}$ case, an inset has been included, showing $P_{2}(t)$ for time $t\in \left[ 0, 4T_{0} \right]$. A gray dashed line marks $t_{r}=d / c_{0}$, indicating that FDTD shows the retardation effect. 
Similar to the vacuum test case, the $\Gamma_{12}$ and $g_{12}$ coefficients are extracted from $P_{1}(t)$, $P_{2}(t)$ with the help of curve fitting process. We compare the estimated values with ground truth obtained from dyadic Green's function, and the results are presented in Fig.~\ref{fig:fig3}(f). Due to the waveguide's confinement effect, the dipole-dipole interaction does not decay quickly with increased distance $d$ but instead oscillates in a sinusoidal manner. 
In conclusion, the proposed FDTD algorithm effectively simulates the excitation transport between two TLSs, capturing the complex dipole-dipole interactions across varying distances and environments. %which is difficult to achieve using semi-classical methods \cite{RDDI_Li_2018}. 
% but also demonstrates significant computational efficiency. Unlike the master equation approach, which requires computing all $\Gamma_{ij}$ and $g_{ij}$ coefficients before starting simulation, for FDTD the explicit calculation of $\Gamma_{ij}$ and $g_{ij}$ coefficients is easily avoided. 

\subsection{Collective behavior of TLS array}
\begin{figure*}[]
\centering\includegraphics[width=1\textwidth]{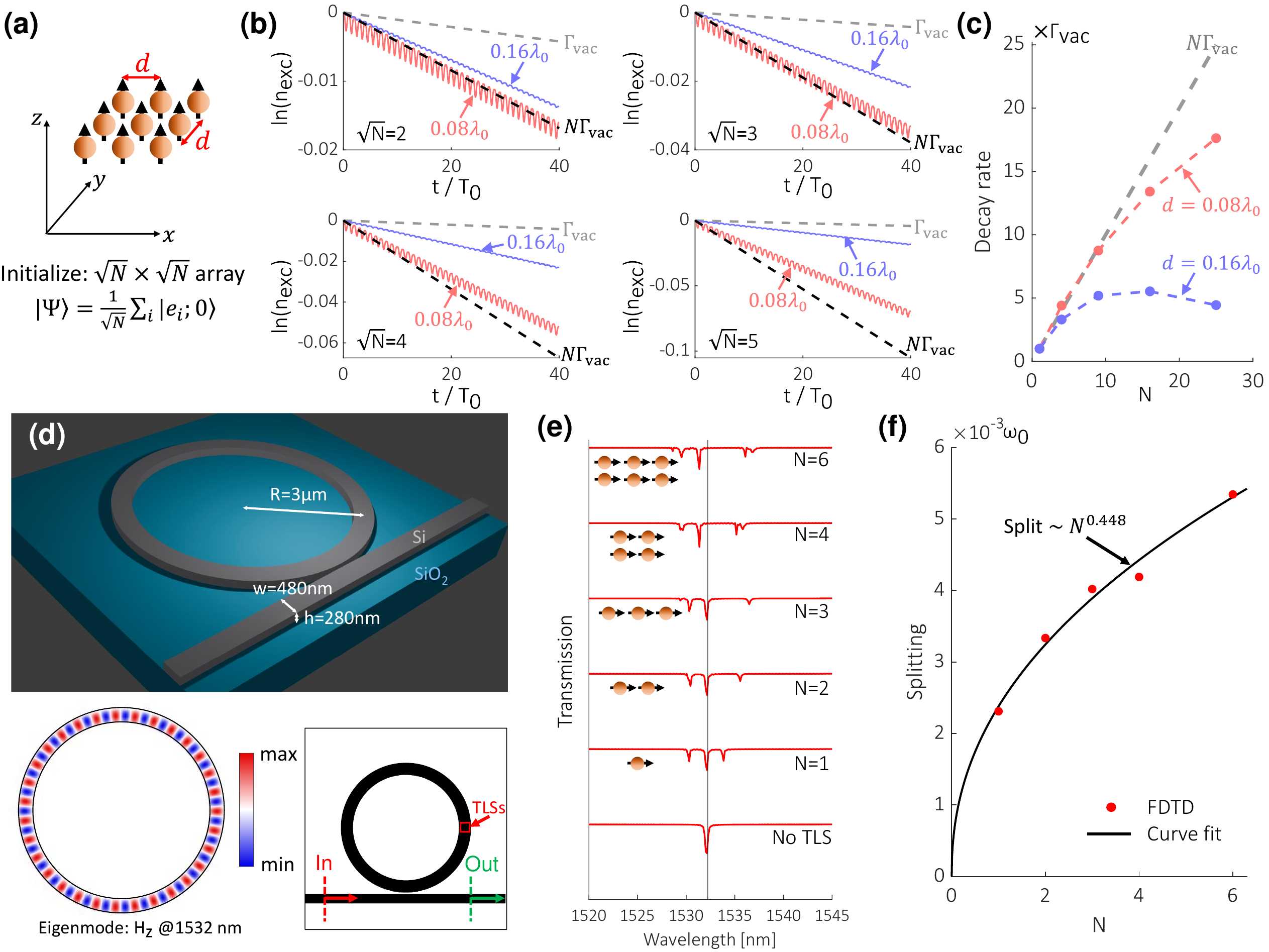}
\caption{3D FDTD simulations involving $N>2$ TLSs. Two examples, namely superradiance and strong-coupling cavity QED, are included. (a) Illustration of the simulated square TLS array. The distance between neighboring TLSs is denoted as $d$. All dipole moments are aligned at $z$ direction. (b) Time-evolution of the total excitation number $n_{\text{exc}}$, under different array sizes and TLS spacings. $\exp(-\Gamma_{\text{vac}}t)$ and $\exp(-N\Gamma_{\text{vac}}t)$ are shown in dashed lines for comparison. Compact arrays show decay rates that are very close too $N\Gamma_{\text{vac}}$, while more extended arrays decay slower. (c) The corresponding decay rates extracted through curve fitting. (d) Illustration of the 3D ring resonator used in cavity QED simulation. The first inset plots the $H_{z}$ distribution of one eigenmode. The second inset shows the input port, output port, as well as the position of TLSs. (e) The transmission spectra when $N$ TLSs are strongly coupled to the ring resonator. Here $N\in \{0, 1, 2, 3, 4, 6\}$. Mode splitting becomes larger when more TLSs are coupled to the resonator. Notice that when $N\geqslant 3$ more than $3$ dips exist in the transmission curve. (f) The relationship between measured Rabi splitting and number $N$. Curve fitting shows that the splitting is proportional to $N^{0.448}$, which increases slightly slower compared to the ideal $\sqrt{N}$ scaling predicted by single-mode cavity QED.}
\label{fig:fig4}
\end{figure*}
In this section, we demonstrate the scalability of our proposed algorithm by simulating the collective behavior of $N>2$ TLSs within a 3D domain. 
We select two representative scenarios: the Dicke superradiance in an ordered TLS array, and Rabi splitting induced by coupling multiple TLSs with a ring resonator. 
To the best of our knowledge, this study is the first to correctly incorporate multiple isolated TLSs into a 3D FDTD simulation. Many previous studies have focused on ensembles of TLSs, allowing them to safely ignore self-interaction \cite{ultrafast_pulse_1995, mbsolve_2021, time_discretize_Maxwell_Bloch_2003, FDTD_ultrashort_pulse_OL_2003, Maxwell_Bloch_PCSEL_Hughes_2017, model_laser_MEEP_2020}; however, some papers that claim to simulate isolated TLS have overlooked self-interaction, leading to results that are qualitatively incorrect \cite{Sha_Maxwell_Schrodinger, Maxwell_Schrodinger_control_pulse_2015, FDTD_TLS_PhC_cavity_Japan_2011, Maxwell_Schrodinger_Ryu_2016, chiral_modeling_2023}. While only a few studies have successfully integrated isolated TLS into FDTD with self-interactions correctly excluded, simulations involving multiple TLSs are still missing \cite{multiscale_Maxwell_Schrodinger_2009, Self_interaction_Maxwell_Liouville, self_consistent_Lorentz_Hughes_2017, nonlinear_Maxwell_Schrodinger_2009}. 
% Paper that sample trajectories
% \cite{Ehrenfest_R_2019, quasiclassical_cavity_QED_Li_2020}
% In the subsequent sections, we demonstrate that with TLSs incorporated, FDTD not only validates known phenomena, but also reveals phenomena that may have been overlooked in prior research. This is because the proposed FDTD method minimizes reliance on commonly used approximations, including the single-mode approximation, rotating-wave approximation (RWA), and Markov approximation. 

% \subsection{Superradiance of ordered TLS array}
We first consider the superradiance phenomenon of an ordered TLS array in vacuum \cite{universality_Dicke_2022, many_Dicke_PRL_2022}. As is well known, TLSs in close proximity to each other tend to synchronize as they decay, resulting in a fast decay rate that is $N$ times larger than that of a single TLS \cite{Scully_timing_2006, scully2009super, Haroche_superradiance_1982}. Such superradiance phenomenon was first predicted by R. Dicke in 1954 \cite{Dicke_1954}, and has already been observed in a wide range of experimental systems \cite{Dicke_observe_in_cavity_2014, tailor_superradiance_2022, Yan_superradiant_cavity_PRL_2023}. However, numerical simulation of such phenomenon based on FDTD are rarely reported \cite{quasiclassical_cavity_QED_Li_2020}. 
The simulation setup is shown schematically in Fig.~\ref{fig:fig4}(a). A square array consisting of $N$ TLSs is positioned at the $z=0$ plane. All TLSs feature a dipole moment oriented in the $z$ direction, each with a magnitude of $0.002$. The distance between neighboring TLSs is denoted as $d$, and in our simulations two different distances $d/\lambda_{0} \in \{ 0.08, 0.16 \}$ have been tested. 
The initial quantum state of the system is set as $| \Psi(t=0) \rangle = \frac{1}{\sqrt{N}} \sum_{i} | e_{i}, 0 \rangle$. We record the total number of excitations $n_{\text{exc}}$, defined as the sum of the excited probabilities $|b_{i}|^{2}$. 
The time-evolution of logarithm $\ln (n_{\text{exc}})$ is plotted in Fig.~\ref{fig:fig4}(b). 
For comparison, the exponential decay curve of an isolated TLS, $\exp(-\Gamma_{\text{vac}}t)$, is shown in gray dashed lines. Additionally, the black dashed line showing $\exp(-N\Gamma_{\text{vac}}t)$ represents the ideal case of superradiance, where all TLSs are in close proximity and oscillate in-phase. 
Four different array sizes ($N=2\times 2$, $3\times 3$, $4\times 4$ and $5\times 5$) have been examined. %As previously mentioned, our FDTD does not rely on RWA, leading to fast oscillations in the $n_{\text{exc}}$ curve. 
It can be concluded that when the array is confined within a region much smaller than the wavelength $\lambda_{0}$, $n_{\text{exc}}$ decays with a rate very close to $N\Gamma_{0}$. Conversely, when the array size is comparable to $\lambda_{0}/2$, TLSs at different locations cannot oscillate perfectly in-phase, resulting in a decay rate smaller than $N\Gamma_{\text{vac}}$. 
Through curve fitting, we have extracted the decay rates from all $n_{\text{exc}}(t)$ curves, with results displayed in Fig.~\ref{fig:fig4}(c). It's evident that when neighboring TLSs are close to each other ($d=0.08\lambda_{0}$ case), the decay rate only deviates from $N\Gamma_{0}$ significantly when $N$ becomes larger than $16$. However, for larger spacing $d=0.16\lambda_{0}$, the decay rate ceases to increase once $N>9$. Our simulation results align with recent studies \cite{universality_Dicke_2022, many_Dicke_PRL_2022}, suggesting that superradiance in an ordered TLS array only occurs when the inter-TLS distance is below a critical value. 

% \subsection{Rabi splitting in cavity}
Next we consider a cavity quantum electrodynamics (cavity QED) system, which is crucial for studying fundamental physics \cite{Kejie_photon_photon_nonlinearity, Ivana_atom_gate_Science_2021, quantum_network_atom_in_cavity_2012, single_atom_cavity_Nature_2014}. To date, numerical simulations relevant to this topic are mostly based on cavity QED theory, typically considering only a few cavity modes (often just a single mode) \cite{Jaynes_1963, cavity_QED_tutorial_2021}. Although these simulation techniques are straightforward to implement, they become inadequate when the photonic environment or the placement of TLSs becomes increasingly complex. 
In this section, we demonstrate the versatility of our algorithm by simulating multiple TLSs strongly coupled to a ring resonator. We consider a silicon ring resonator placed on top of a silica substrate, as depicted in Fig.~\ref{fig:fig4}(d). The ring resonator is coupled to an adjacent silicon waveguide, positioned $80$ nm from the ring. Additional size parameters are provided in Fig.~\ref{fig:fig4}(d). 
The ring resonator features a free spectral range (FSR) of approximately $369$ THz. The magnetic field distribution $H_{z}$ for one eigenmode at $1532$ nm is visualized in Fig.~\ref{fig:fig4}(d). Multiple TLSs, with dipole moments oriented in $x$ direction, are embedded within the ring resonator. 
As shown by the insets of Fig.~\ref{fig:fig4}(e), these TLSs form an array, which is centered at position $(3\mu\text{m}, 0, 0)$. The distance between neighboring TLSs remain fixed as $d=0.08 \lambda_{0}$. 
The TM0 waveguide mode is injected from the input port (indicated in red) on the left side, and the transmission through the system is measured at the output port (indicated in green) on the right side. The corresponding transmission spectra for different number of TLSs $N\in \{ 0, 1, 2, 3, 4, 6 \}$ are depicted in Fig.~\ref{fig:fig4}(e). 
The resonance wavelength $\lambda_{0}$ is highlighted with a vertical line. 
In the absence of TLSs, the transmission spectrum features a single dip. Coupling a single TLS ($N=1$) with the ring resonator results in three dips. This effect arises because the ring resonator supports two degenerate cavity modes: one traveling clockwise (CW) and the other counter-clockwise (CCW). With a dipole moment of $0.01$ oriented in the $x$ direction, the TLS interacts with both CW and CCW modes, leading to the splitting of three distinct modes. This phenomenon, known as vacuum Rabi splitting \cite{vacuum_Rabi_splitting_2004}, indicates that the dipole moment is sufficiently large to achieve the strong-coupling regime. 
As the number of TLSs coupled with the ring resonator increases, the mode splitting becomes larger. The relationship between splitting (measured by subtracting the frequencies of two most distant dips) and the number of TLSs $N$ is plotted in Fig.~\ref{fig:fig4}(f). According to cavity QED theory, when $N$ TLSs are coupled to a single cavity mode, the Rabi splitting should scale as $\sqrt{N}$ \cite{Yan_superradiant_cavity_PRL_2023, mark_fox_quantum_optics_textbook, strong_couple_FP_cavity_PRL_2023}. In our simulations, the relationship between splitting and $N$ has been determined through curve fitting. Our FDTD simulations reveal that the splitting is proportional to $N^{0.448}$, which is slightly lower compared to the theoretical $N^{0.5}$ scaling. This deviation can be attributed to the inter-TLS distance $d=0.08\lambda_{0}$, which results in variations in the coupling constants among different TLSs. 
Notice that for all cases with $N\geqslant 3$, more than 3 dips appear in the transmission spectra. %This observation might seem counter-intuitive, as single-mode cavity QED theory predicts that the number of transmission dips (or peaks) should not increase with $N$. 
We speculate that these additional dips correspond to eigenmodes typically referred to as ``dark states'' \cite{polariton_chem_review_Mandel}. 
These dark states are not entirely ``dark'' when considering the near-field coupling between TLSs \cite{Liu_photon_scatter_OL_2016, Liu_enhance_cross_section_PRA_2017, superradiance_Rydberg_arr_2024, multimode_dark_states_2011}, an interaction that is overlooked in conventional single-mode cavity QED treatments \cite{multimode_lossy_PRL_2023}. %We are currently trying to gain a deeper understanding of these unexpected quantum states. Exploring the potential applications of these dark states would be an intriguing direction for future research. 

\section{Discussion}
% Objective
In this work, we aim to incorporate quantum TLSs into well-developed FDTD simulation framework. This integration provides a methodology to analyze the behavior of multiple TLSs within various photonic environments.  
% Main findings
We pointed out that many existing simulation techniques have failed to correctly exclude primary radiation fields. %resulting in spurious frequency shifts. 
Our modifications are two-fold: theoretically, we restrict our analysis to single-excitation quantum states, which has allowed us to precisely define the current sources for modeling photon emission; numerically, we introduce the TF-IF technique, utilizing auxiliary FDTDs to exclude primary radiation fields. 
% Advantages: capabilities, problems solved
By making the above modifications, we have enabled the simulation of TLS dynamics within complex photonic environments, with spurious self-interactions eliminated. 
% Benchmark
This paper presents several test cases to validate the accuracy of our FDTD algorithm. 
For a single TLS ($N=1$) we have confirmed that our FDTD accurately simulates both photon scattering and spontaneous emission. % N=1
For two TLSs we focus on the excitation transport between two distant TLSs. Notably, our FDTD simulations reproduce the retardation effects due to the finite speed of light, thereby maintaining causality.  % N=2
% Collective behavior
To further demonstrate the scalability of our algorithm, we have included two more examples involving $N>2$ TLSs. Based on the results, our FDTD can correctly simulate superradiance effect of TLS array, as well as vacuum Rabi splitting when multiple TLSs couple strongly with a ring resonator. 

% Summarize briefly
In summary, the proposed FDTD algorithm accurately captures a variety of phenomena that are of great interest to researchers. 
Additionally, FDTD can predict novel behaviors that have been overlooked in past research works, due to the use of various simplifications. 
Our FDTD is scalable when the number of TLSs increases, because the overhead is caused by introducing auxiliary FDTDs, which are much smaller in size compared with the main FDTD. 
The algorithm has been implemented using CUDA C++, with the code made publicly available on \href{https://github.com/zhouqingyi616/FDTD_w_TLS}{GitHub}. 
We are confident that our modified FDTD will serve as a powerful tool, bridging the gap between theoretical predictions and experimental verifications. %: theorists can test their hypotheses without physical experiments, while experimentalists gain a more realistic simulation platform that can predict outcomes prior to laboratory experiments. 
% Limitations: challenges
Despite these advantages, our FDTD still relies on the assumption of single-excitation states. Investigating whether similar numerical techniques could be adapted to systems containing multiple entangled TLSs and photons would be a promising direction for future research.
% Future directions: enhance, new applications
Currently, we are integrating this algorithm into Tidy3d \cite{Tidy3D}, a well-regarded FDTD simulation platform, with the aim of making it accessible to the research community shortly. 
As part of our long-term objectives, we plan to enhance the proposed technique to encompass more complex quantum emitters, including atoms and molecules with multiple energy levels. 
%A simulation technique that can accurately simulate the dynamics of molecular interactions inside complex environments is highly desired. 
We anticipate that the development of such tool will significantly advance research and enhance our understanding across a wide range of fields, including quantum physics, chemistry, as well as biology. 

\section{Methods}
\subsection{Procedure of the proposed FDTD}
Here the main steps of our proposed FDTD algorithm are summarized. For the sake of simplicity, we mark the fields calculated by the $i$-th auxiliary FDTD with footnote $i$. For example, $\vec{H}_{i}^{n}$ corresponds to the $\vec{H}$ field calculated by the $i$-th auxiliary FDTD at time $n\Delta t$. On the other hand, the fields calculated by the main FDTD do not have footnote. 
At the $n$-th time step, suppose we have already obtained fields $\vec{E}^{n-\frac{1}{2}}$ and $\vec{H}^{n}$, as well as $\vec{E}_{i}^{n-\frac{1}{2}}$ and $\vec{H}_{i}^{n}$ for all auxiliary FDTDs. 
The modified FDTD contains the following $9$ steps: \\
(1) For all $N$ auxiliary FDTDs: based on $\vec{H}_{i}^{n}$ values on box boundary $\partial \Omega_{i}$, calculate the fictitious current source $\vec{J}^{n}_{i, \text{rad}} \sim \hat{n} \times \vec{H}_{i}^{n}$, which will be used in the main FDTD; \\
(2) Main FDTD: Calculate $\vec{E}^{n+1/2}$ based on $\vec{E}^{n-1/2}$ and all $\vec{J}^{n}_{i, \text{rad}}$'s; \\
(3) For all $N$ auxiliary FDTDs: calculate $\vec{E}_{i}^{n+\frac{1}{2}}$ based on $\vec{E}_{i}^{n-\frac{1}{2}}$ and $\vec{J}_{i}^{n}$; \\
(4) For all $N$ auxiliary FDTDs: based on $\vec{E}_{i}^{n+\frac{1}{2}}$ values on box boundary $\partial \Omega_{i}$, calculate the fictitious magnetic current source $\vec{M}^{n+\frac{1}{2}}_{i, \text{rad}} \sim -\hat{n} \times \vec{E}_{i}^{n}$, which will be used in the main FDTD; \\
(5) Main FDTD: Calculate $\vec{H}^{n+1}$ based on $\vec{H}^{n}$ and all $\vec{M}^{n+\frac{1}{2}}_{i, \text{rad}}$'s; \\
(6) For all $N$ auxiliary FDTDs: calculate $\vec{H}_{i}^{n+1}$ based on $\vec{H}_{i}^{n}$; \\
(7) For all $N$ TLSs: update $b_{i}^{n}$ to $b_{i}^{n+1}$ based on $\vec{E}^{n+\frac{1}{2}}(\vec{r}_{i})$; \\
(8) For all $N$ TLSs: based on coefficient $b_{i}^{n}$, calculate current sources $\vec{J}_{i}^{n}$, which will be used in the auxiliary FDTD; \\
(9) Repeat the above steps until the stopping criteria is satisfied. 
The details of the above procedure is shown schematically in Fig.~S{?}. The implementation details related to 3D FDTD (involving step 2, 3, 5, 6) are provided in Supporting Information, Note S{?}. The calculation of fictitious current sources $\vec{J}^{n}_{i, \text{rad}}$ and $\vec{M}^{n+\frac{1}{2}}_{i, \text{rad}}$ (involving step 1, 4) is described in Supporting Information, Note S{?}. 

\subsection{Numerical solver for TLS's dynamics}
Here we provide details for the implementation of TLS's time-evolution. The differential equations used for 2 baseline semi-classical methods are also provided. 
For the $i$-th TLS located at position $\vec{r}_{i}$, we first sample the electric field $\vec{E}^{n+\frac{1}{2}}(\vec{r}_{i})$ at its location from the main FDTD. This value is then utilized to drive the TLS, by doing a time-marching of the following equation:
\begin{equation}
    \frac{db_{i}}{dt} = (-i\omega_{0} - \frac{\Gamma_{\text{vac}}}{2}) b_{i} + \frac{i}{\hbar} \vec{d}_{i} \cdot \vec{E}(\vec{r}_{i}, t), 
\end{equation}
where $\Gamma_{\text{vac}}$ stands for the TLS's spontaneous decay rate inside vacuum. This decay term is included due to the fact that the primary radiation field has been excluded in $\vec{E}^{n+\frac{1}{2}}(\vec{r}_{i})$. 
As for Maxwell-Schr$\ddot{\text{o}}$dinger equations, the quantum state of the $i$-th TLS can be represented using 2 coefficients: $|\Psi_{i}(t)\rangle = c_{i,g}(t) |g\rangle + c_{i,e}(t) |e\rangle$. The time-evolution of these coefficients follow
\begin{equation}
\begin{aligned}
& \frac{\partial c_{i,e}}{\partial t} = -i\omega_{0} c_{i,e} + \frac{i}{\hbar} \vec{d}_{i} \cdot \vec{E}(\vec{r}_{i}, t) c_{i,g}, \\
& \frac{\partial c_{i,g}}{\partial t} = \frac{i}{\hbar} \vec{d}_{i} \cdot \vec{E}(\vec{r}_{i}, t) c_{i,e}. 
\end{aligned}
\end{equation}
% By looking at the above equations, we can notice that if a single TLS starts from excited state with $c_{i, g}(t=0)=0$, its coherence $c_{i, e}^{*} c_{i, g}$ will remain zero for $t>0$, therefore cannot produce nonzero field. 
For Maxwell-Bloch equations, the time-evolution of the $i$-th TLS involves elements of density matrix $\hat{\rho}_{i}$:
\begin{equation}
\begin{aligned}
& \frac{\partial \rho_{i, ee}}{\partial t} = \frac{i}{\hbar} \vec{d}_{i} \cdot \vec{E}(\vec{r}_{i}, t) \cdot (\rho^{*}_{i, eg} - \rho_{i, eg}) - \Gamma_{\text{vac}} \rho_{i, ee}, \\
& \frac{\partial \rho_{i, eg}}{\partial t} = (-i\omega_{0} - \frac{\Gamma_{\text{vac}}}{2})\rho_{i, eg} + \frac{i}{\hbar} \vec{d}_{i} \cdot \vec{E}(\vec{r}_{i}, t) \cdot (1-2\rho_{i, ee}). 
\end{aligned}
\end{equation}
%Similarly, if a single TLS starts from excited state with $\rho_{i, eg}(t=0)=0$, its coherence $\rho_{i, eg}$ will remain zero for $t>0$. Even though the population $\rho_{i, ee}$ decays exponentially, no electromagnetic fields can be excited when using semi-classical simulation techniques. 
The above sets of equations can be solved using any type of differential equation solver. In this paper, instead of using Euler method (which often leads to divergence), we apply the 4-th order Runge-Kutta method, with a time step of $\Delta t / 5$ to ensure accuracy. 

% \subsection{References} % Ref.~\citenum{Mena2000}

\begin{acknowledgement}
The authors thank Prof. Jennifer T. Choy, Prof. Shimon Kolkowitz, Dr. Momchil Minkov, as well as Dr. Boyuan Liu for many helpful discissions. 
This work was supported by the National Science Foundation’s QLCI-CI: Hybrid Quantum Architectures and Networks, as well as the Army Research Office through a Multidisciplinary University Research Initiative program (Grant No. W911NF-22-2-0111).

\end{acknowledgement}

\begin{suppinfo}
The following files are available free of charge.
\begin{itemize}
  \item FDTD\_TLS\_paper\_SI.pdf: Detailed derivation of equations of motion, for both the electromagnetic fields and the TLSs; Implementation details of the modified FDTD, with TF-IF technique incorporated. 
\end{itemize}

\end{suppinfo}

\bibliography{FDTD+TLS}

\end{document}